\documentclass[prl, aps, superscriptaddress, twocolumn, balancelastpage, 
longbibliography]{revtex4-1}
\usepackage{color}
\usepackage{natbib}
\setcitestyle{numbers}
\setcitestyle{square}
\usepackage{url}

\usepackage[colorlinks,bookmarks=false,citecolor=blue,linkcolor=blue,urlcolor=blue]{hyperref}
\usepackage[all]{hypcap}   % let hyperlinks correctly point to figures rather 
\usepackage{amsmath,amssymb}
\usepackage{graphicx}
\graphicspath{{figures/}{/}}
\usepackage{dcolumn}
\usepackage{bm}
\usepackage[capitalise]{cleveref}
\crefname{section}{Sec.}{Secs.}
\Crefname{section}{Section}{Sections}
\usepackage{bbold}
\usepackage{soul}

\usepackage{titlesec}

\usepackage{verbatim}
\usepackage{color}
\usepackage{soul}
\usepackage{placeins}  % for FloatBarrier
\usepackage{flafter}  

\usepackage{dsfont}

\newcommand{\tr}{{\text{Tr}}}

\newcommand{\unity}{1\!\!1}

\makeatletter
\let\Hy@backout\@gobble
\makeatother

\usepackage{multirow}
\usepackage{array}
\usepackage{tabularx}
\graphicspath{{../figures/}}
\newcommand{\bgl}{Bogoliubov transformation}
\makeatletter
\newcommand*{\rom}[1]{\expandafter\@slowromancap\romannumeral #1@}
\makeatother

\begin{document}
	\title{Quantum criticality and universality in stationary 
		state of long-range Kitaev model}
	\author{Akash Mitra}
	\affiliation{Variable Energy Cyclotron Centre, 1/AF Bidhannagar, 
		Kolkata 
		700064, India}
	\affiliation{Homi Bhabha National Institute, Training School 
		Complex, 
		Anushaktinagar, Mumbai - 400094, India}
	\author{Sanku Paul}
	\affiliation{Department of Physics and Complex Systems, S.N. Bose 
		National 
		Centre for Basic Sciences, Kolkata 700106, India}
	\author{Shashi C. L. Srivastava}
	\affiliation{Variable Energy Cyclotron Centre, 1/AF Bidhannagar, 
		Kolkata 
		700064, India}
	\affiliation{Homi Bhabha National Institute, Training School 
		Complex, 
		Anushaktinagar, Mumbai - 400094, India}
	\begin{abstract}
		We investigate the signature of  quantum criticality 
		in the long-time stationary state of the long-range Kitaev chain by 
		performing various quench protocols. In this model, the 
		pairing interaction decays with distance according to a power 
		law with exponent $\alpha$. Using quantum information-theoretic 
		measures, such as mutual information and logarithmic negativity, we 
		show that, irrespective of the values of $\alpha$, critical-to-critical 
		quench displays quantum criticality even in the stationary state. 
		Remarkably, in the presence of long-range pairing 
		interactions, where fermionic correlators decay algebraically 
		even at noncritical points, the signature of quantum criticality 
		persists in the stationary state. Furthermore, the effective central 
		charge, calculated  from both mutual information and 
		logarithmic negativity of the stationary state following a 
		critical-to-critical quench, agrees with the central 
		charge of the corresponding ground states for both $\alpha = 0$ and 
		$\alpha = 2$. Therefore, information of the universality class can be 
		inferred from the stationary state.

	\end{abstract}
	\maketitle
	\twocolumngrid
	\section{Introduction}
	
	Understanding quantum phase transitions (QPTs) has been a central topic of 
	interest over the past few decades due to its significant implications for 
	understanding the collective behavior of many-body systems \cite{Sachdev_2011}. 
	QPTs have been both theoretically and experimentally studied extensively in 
	various systems, including quantum spin chains 
	\cite{Osborne_2002,Peng_2005,Raoul_2008,Kargarian_2008,Maziero_2010,Ma_2011,Sachdev_2011,Li_2011,Rams_2012,Yao_2012,Cheng_2012,Blanc_2018},
	Bose-Einstein condensates 
	\cite{Jaksch_1998,Greiner_2002,Capogrosso_2007,Donner_2007,Spielman_2007,Zhou_2011,Rubeni_2012,Zhang_2012},
	and strongly correlated electron systems 
	\cite{Si_2001,Vojta_2003,ALET_2006,chern_2014,Paschen_2021}. 
	QPT occurs at specific values of the parameter(s) present 
	in the Hamiltonian, called critical points, where the system exhibits scale 
	invariance. A universal behavior ensues due to the scale invariance, which can 
	be classified into certain universality classes based on the space dimension 
	and symmetry of the order 
	parameter \cite{Stanley_1999,Cardy_1996,kadanoff_1971,Pelissetto_2002}.
	This universality is studied using conformal field theories, and the 
	corresponding  central 
	charge value is commonly used to identify the underlying 
	universality class \cite{BELAVIN_1984,Affleck_1986,Francesco_1997}.
	For instance, systems belonging to the Ising universality class are 
	described by a minimal model with a central charge $c=1/2$, while the Luttinger 
	liquid universality class corresponds to a conformal bosonic theory with 
	$c=1$ \cite{Blote_1986,Ginsparg_1988}. Note that at 
	quantum critical points, the development of 
	long-range correlations leads to an algebraic decay of correlators over 
	distance, in contrast to the exponential decay typically observed at 
	noncritical points in Hamiltonians with only short-range interactions 
	\cite{STANLEY_1993,Hastings_2006,brandao_2013,brandao_2015}. 
	This is often used to identify critical points.

	While the existence of quantum criticality and long-range correlations in the 
	ground states of many-body quantum systems are widely studied, their presence 
	in 
	stationary states is not yet well explored. In Ref. \cite{Eisler_2014}, it 
	is 
	shown that when two sides of a one-dimensional (1D) noninteracting fermionic 
	chain with nearest-neighbor hopping are prepared at different temperatures, 
	the mutual information scales 
	logarithmically with subsystem size in the steady state, indicating the 
	presence of long-range correlations. Another example of the survival of 
	quantum criticality in the long-time stationary state is shown in the 
	anisotropic $XY$ chain under a sudden quench protocol \cite{Sanku_2024}. 
	Using quantum information measures such as mutual 
	information and logarithmic (log-) negativity, it is shown that when both 
	the 
	pre- and postquench parameters of the Hamiltonian are set at the 
	critical point,  both mutual information and log-negativity exhibit 
	a peak, indicating the presence of criticality in the stationary 
	state. In contrast, the peak in mutual information and 
	log-negativity vanishes for other quench protocols when either/both 
	pre- and 
	postquench parameters  differ from critical values. 
	The signature of quantum criticality in the stationary state is attributed 
	to a 
	change in correlation pattern from exponential to algebraic decay, 
	which was captured by mutual information and log-negativity.
	
	A natural question follows: What about the scenario when correlators decay 
	algebraically, even at noncritical points? Long-range interacting systems 
	provide an ideal platform to explore this question as fermionic correlators 
	have been shown to decay algebraically over distance, even when the 
	Hamiltonian is gapped \cite{Koffel_2012, Vodola_2015, Eisert_2010, 
	Vodola_2014}, 
	i.e., at noncritical points. These long-range interacting systems 
	have been 
	studied in different contexts both theoretically and experimentally in 
	atomic, 
	molecular, and optical lattice systems \cite{Saffman_2010, 
		Lahaye_2009, Firstenberg_2013, Ritsch_2013, Yan_2013, Schau__2012, 
		Lu_2012, Gopalakrishnan_2011, 
		Kim_2009,Britton_2012,Schneider_2012,Islam_2013,Richerme_2014,
		Jurcevic_2014,Douglas_2015,Landig_2015,Landig_2016}.
	We take the 1D long-range Kitaev model with a pairing term that decays with 
	distance as a power law $\propto l^{-\alpha}$, with $\alpha$ being the 
	exponent \cite{Vodola_2014}. From an experimental perspective, 
	this model is particularly relevant as it is closely related to the Ising model 
	with tunable long-range interactions, which can be experimentally realized 
	using trapped ion setups \cite{Kim_2009,Britton_2012, Schneider_2012, 
		Bermudez_2013, Jurcevic_2014, Richerme_2014}. In the ground state, this model 
	undergoes an exotic transition from the Ising-type universality class, observed 
	for $\alpha>3/2$, to a Luttinger liquid universality class at $\alpha=0$. 
	
	In this paper, we investigate whether it is possible to capture quantum 
	criticality in the stationary state when fermionic correlators decay 
	algebraically even at noncritical points. Additionally, we examine  
	whether the stationary state can be 
	described by the same universality class as the ground state by analyzing the 
	central charge obtained via scaling of mutual information and log-negativity 
	for various quench protocols. Specifically, we consider three values of 
	$\alpha$ 
	corresponding to different universality classes observed in the ground state: 
	first, corresponding to the Ising universality class ($\alpha=2$); second, 
	the Luttinger liquid universality class ($\alpha=0$); and a third value 
	($\alpha=1$) where there is no universality \cite{Vodola_2014}. 
	{To understand the tripartite information in the postquench 
	stationary state and whether signatures of criticality will manifest, we 
	study tripartite mutual information.}

	\section{The model and ground state phase transition} \label{Sys}
	The Hamiltonian of the $1$D long-range Kitaev model (LRK) for a lattice site of 
	length $N$ is expressed as \cite{Vodola_2014}
	\begin{equation}\label{eq:LRK_ham}
		\begin{aligned}
			H_{\rm LRK}= &\sum_{j=1}^{N} \left[ -t 
			\left(f_j^{\dagger}f_{j+1}+f_{j+1}^{\dagger}f_j\right)-\mu 
			\left(f_j^{\dagger}f_j-\frac{1}{2}\right)\right.\\
			&\left. +\frac{\Delta}{2}\sum_{l=1}^{N-1}\frac{1}{l^{\alpha}}\left( 
			f_{j+l}^{\dagger}f_j^{\dagger}-f_{j+l}f_j\right)\right]\,
		\end{aligned} 
		%\label{eq:LRK}
	\end{equation}
	where $f_j(f_j^{\dagger})$ is the fermionic annihilation (creation) operator at 
	site $j$, satisfying the canonical anticommutation relations $\{f_i, 
	f_j^{\dagger}\}=\delta_{ij}$, $\{f_i,f_j\}=\{f_i^{\dagger},f_j^{\dagger}\}=0$. 
	The parameters $t$ and $\mu$ represent the tunneling rate between two 
	neighboring sites and the chemical potential, respectively. We set $2t=1$ 
	in the rest of the paper. The parameter, $ \Delta$ denotes 
	the strength of the fermion $p$-wave pairing interaction, while its range is 
	governed by the exponent $\alpha\in 
	[0,\infty)$. The two limits, i.e., $\alpha=0$ and $\alpha \to \infty$, correspond 
	to 
	all-to-all interaction with equal strength and nearest-neighbor 
	interaction, respectively. We consider the antiperiodic boundary condition 
	throughout 
	the paper. The Hamiltonian in \cref{eq:LRK_ham} is exactly solvable. To see 
	this, we first 
	perform a Fourier transformation to transform \cref{eq:LRK_ham} in the momentum 
	space as
	\begin{equation}\label{eq:Ham_fer}
		\begin{aligned} 
			H_{\rm LRK}=\frac{1}{2}&\sum_{n=0}^{N-1}
			\begin{bmatrix}
				f_{k_n}^\dagger & f_{N-k_n}
			\end{bmatrix}\\
			&\begin{bmatrix}
				-\left(\mu+\cos k_n\right) & i \Delta g_\alpha(k_n) \\
				-i \Delta g_\alpha(k_n) & \left(\mu+\cos k_n\right)
			\end{bmatrix}
			\begin{bmatrix}
				f_{k_n} \\ f_{N-k_n}^\dagger
			\end{bmatrix}\,,
		\end{aligned}
	\end{equation}
	where $g_\alpha(k)=\sum_{l=1}^{N-1} \frac{\sin (kl)}{l^\alpha}$ and 
	$k_n=\frac{2\pi}{N}\left(n+1/2\right)$. In the large-$N$ limit, $g_\alpha(k)$ takes the form 
	\begin{equation}
		g_\alpha(k)=-\frac{i}{2} 
		\left[\text{Li}_\alpha\left(e^{ik}\right)-\text{Li}_\alpha\left(e^{-ik}\right)\right]\,.
	\end{equation}
	The function $\rm Li_{\alpha}(z)$ represents the polylogarithm of the complex 
	variable $z$ of order $\alpha$. The Hamiltonian in \cref{eq:Ham_fer} can 
	further be cast to a diagonal form by performing the following \bgl:
	\begin{align}\label{eq:transform_Bogolyubov}
		\begin{bmatrix}
			f_{k_n} \\
			f_{N-k_n}^\dagger
		\end{bmatrix}= \begin{bmatrix}\cos \theta_{k_n} & -i\sin \theta_{k_n} \\
			-i \sin \theta_{k_n} & \cos \theta_{k_n}
		\end{bmatrix}\begin{bmatrix}
			\eta_{k_n} \\
			\eta_{N-k_n}^\dagger
		\end{bmatrix},
	\end{align}
	with $\theta$ being the Bogoliubov angle, and defined as
	\begin{equation}\label{eq:bg_angle}
		\tan(2\theta_{k_n})=\frac{\Delta g_\alpha(k_n)}{\mu+\cos k_n}.
	\end{equation}
	The diagonal form of the Hamiltonian in \cref{eq:LRK_ham} in the basis of 
	Bogoliubov fermions $\eta$ then becomes 
	\begin{equation}\label{eq:Ham_int}
		H_{\rm LRK}=\sum_{n=0}^{N-1} \lambda_\alpha(k_n) \left(\eta_{k_n}^\dagger 
		\eta_{k_n}-\frac{1}{2}\right),
	\end{equation}
	where the dispersion relation is
	\begin{equation}\label{eq:Ek_exp}
		\lambda_\alpha(k_n)=\sqrt{ \left(\mu+\cos k_n\right)^2+\left(\Delta 
			g_\alpha(k_n)\right)^2}.
	\end{equation}
	Equation (\ref{eq:Ham_int}) signifies that each mode is independent, implying the integrability of the system. In the limit $\alpha \to \infty$, 
	only nearest-neighbor terms contribute to 
	the sum in $g_\alpha(k)$, resulting in 
	$g_{\infty}(k)=\sin(k)$. The Hamiltonian in \cref{eq:LRK_ham} in this limit 
	coincides with the Hamiltonian of the $XY$ model obtained via the Jordan-Wigner 
	transformation. For the $XY$ model, a ground-state phase transition occurs 
	from the gapped ferromagnetic ordered phase to the gapped paramagnetic 
	disordered 
	phase with increasing the chemical potential $|\mu|$, with the two critical 
	gapless points at $\mu = \pm 1$  \cite{Vodola_2014}. 
	The spectrum in \cref{eq:Ek_exp} continues to be gapless at $\mu=\pm 1$, even 
	with the smaller values of $\alpha$, i.e., in the presence of 
	long-range 
	interactions as long as $\alpha > 1$. However, the phase diagram changes for 
	$\alpha<1$. This regime is referred to as ``strong" long-range regime since 
	$\alpha<d$, with $d$ representing the dimension of the system. In comparison 
	to $\alpha > 1$, one significant change is that $\mu=-1$ is no longer a 
	critical point, and the phase diagram is no longer symmetric across the line 
	$\mu=0$ \cite{Vodola_2014,Vodola_2015}.
	\begin{table}[h]
		\caption{Summary of the effective central charge $c_{\rm eff}$ for 
		different values 
			of $\alpha$ and 
			$\mu$ for the ground state of the long-range Kitaev model 
			\cite{Vodola_2014,Vodola_2015, Sowiski_2014, Ares_2015}.}\label{tab:ceff_gs}
		\begin{tabular}{|c|c|c|c|}
			\hline 
			~ & $\alpha=0$ & $0<\alpha \leq 1$ & $\alpha \to \infty$\\ \hline 
			$\mu = 1$ & 1 & $c_\text{eff}(\alpha, \Delta)\neq 0$ & $\dfrac{1}{2}$\\ [2mm]
			$\mu \neq \pm 1$ & $\dfrac{1}{2}$ & $c_\text{eff}(\alpha, \Delta) \neq 
			0$ &  
			0\\ [2mm]
			\hline
		\end{tabular}
	\end{table}
	A good probe to identify these critical points is the von Neumann 
	entropy $S_{vN}$, expressed as $S_{vN}=-\tr \left(\rho_L \ln \rho_L\right)$, 
	where $\rho_L$ is the reduced density matrix of the subsystem of length $L$.
	At these critical points, where the spectral gap closes, $S_{vN}$ for 
	short-range interacting system scales as
	\cite{Holzhey_1994,Vidal_2003,Calabrese_2004,Calabrese_2009}
	\begin{equation}\label{eq:entropy_subsystem}
		S_{vN} \approx \frac{c}{3} \ln \left[\frac{N}{\pi} \sin \left(\frac{\pi 
			L}{N}\right)\right]+c^\prime,
	\end{equation}
	where $c$ is the central charge of the underlying conformal field theory 
	(CFT), $N$ is the total system size, and $c^\prime$ is a nonuniversal constant. In fact, 
	\cref{eq:entropy_subsystem} is also valid for all values of $\alpha$ and 
	$\mu$, but 
	the central charge $c$ is now replaced with an effective central charge $c_{\rm 
		eff}$ \cite{Vodola_2014,Sowiski_2014}. 
	The values of the $c_{\rm eff}$ for the ground state are summarized in 
	Table \ref{tab:ceff_gs}. It is worth mentioning that $\alpha=0$ signifies 
	all-to-all 
	coupling between the sites and falls under the Tomonaga-Luttinger liquid 
	universality class \cite{Cazalilla_2011}.

	\section{Methods}\label{Quenp}
	In the last section, we summarized criticality seen in the ground state of the LRK 
	model and (effective) central charge of the underlying CFT. A natural 
	question then arises: Does this criticality 
	survive 
	when the system undergoes out-of-equilibrium dynamics? In other words, in a 
	stationary state, can we find any signature of quantum criticality? A common 
	way to achieve the nonequilibrium dynamics 
	is through sudden quench where  the local or global parameters of the system 
	are suddenly changed \cite{Dorner_2012}. As a consequence, the ground state of 
	the prequench Hamiltonian is no longer a ground state of the 
	postquench 
	Hamiltonian, but a superposition of its eigenstates. The long-time stationary 
	state of the LRK model under sudden quench is the system of 
	interest.  To probe the criticality, much like the ground state, entanglement 
	entropy would be our first choice. However, as the long-time stationary 
	state 
	typically involves highly excited states,  $S_{vN}$ exhibits volume-law 
	scaling eclipsing the logarithmic dependence on subsystem size which may be 
	present due to criticality. So, to extract 
	these logarithmic correlations, we consider quantum 
	information-theoretic measures such as mutual information and 
	log-negativity. For the
	ground state, mutual information has already been established to capture the 
	criticality in fermionic systems \cite{Lepori_2022}.

	\textit{Mutual information}. The mutual information between two subsystems 
	$A_1$ and $A_2$ is defined as
	\begin{equation}\label{eq:MI_def}
		I_{A_1:A_2}=S_{vN}^{A_1}+S_{vN}^{A_2}-S_{vN}^{A_1 \cup A_2},
	\end{equation}
	where $S_{vN}^{A_1} (S_{vN}^{A_2})$ is the von Neumann entropy of the subsystem 
	$A_1 (A_2)$ and $S_{vN}^{A_1 \cup A_2}$ is the von Neumann entropy of $A_1 \cup 
	A_2$. Throughout the paper, we consider the subsystem sizes $|A_1|=|A_2|=L$. 
	The mutual information between two subsystems measures the total amount of 
	information (both classical and quantum) that one subsystem contains about the 
	other \cite{Groisman_2005}. Furthermore, if $S_{vN}$ of the long-time 
	stationary state contains both the volume and logarithmic terms, the mutual 
	information should scale as $ I_{A_1:A_2} \sim b \ln L$. Thus, if the 
	long-range correlations are present in the stationary state, then the 
	mutual information should behave as 
	\cite{Dai_2021,Stephan_2014,Alcarez_2014,Wilms_2012,Eisler_2014,Sanku_2024}
	\begin{equation}
		I_{A_1:A_2} \sim \frac{c_{\rm eff}^{I}}{3} \ln L +\rm const,
	\end{equation}
	where we define $c_{\rm eff}^{I}$ as the effective central charge of the 
	stationary state 
	extracted using mutual information. In this way, mutual information can capture 
	the logarithmic correlation that is hidden in $S_{vN}$. However, 
	whether these 
	correlations are truly quantum in nature remains to be investigated. In this 
	context, log-negativity serves as a useful measure.

	\textit{Log-negativity.}: Similar to mutual information, log-negativity, an 
	entanglement monotone, also quantifies the nonlocal 
	correlation between two subsystems. But unlike mutual information, it captures 
	only quantum correlations 
	\cite{Plenio_2005,Vidal_2002,Eisert_1999,Horodecki_1996}. It is defined as 
	\cite{Vidal_2002,Plenio_2005}
	\begin{equation}\label{eq:logneg_defn}
		\xi_{A_1:A_2} \equiv \ln ||\rho_A^{T_2}||=\ln \tr |\rho_A^{T_2}|,
	\end{equation}
	where $||.||$ denotes the trace norm, and $\rho_A^{T_2}$ signifies the partial 
	transposition of the reduced density matrix of $A \equiv (A_1 \cup 
	A_2)$. It is 
	worth mentioning that 
	for a CFT, 
	the expression of log-negativity for two adjacent subsystems each of length 
	$L_1$ and $L_2$, respectively, %in an infinite system 
	is shown to be \cite{Calabrese_2012}
	\begin{equation}
		\xi_{A_1: A_2}=\frac{c}{4} \ln \left(\frac{L_1 L_2}{L_1+L_2}\right),
	\end{equation}
	where $c$ is the central charge. For the $XY$ model, during a 
	critical-to-critical quench, the log-negativity of the long-time stationary 
	state scales as $\xi_{A_1: A_2}= \frac{1}{8} \ln L$, where 
	$L_1=L_2=L$ \cite{Sanku_2024}. The logarithmic divergence of $\xi_{A_1: A_2}$ 
	is consistent with the critical 
	ground state, as the $XY$ model belongs to the Ising universality class 
	with $c=1/2$. Based on this, we assume that if there is any 
	signature of long-range quantum correlations in the long time stationary state, 
	the log-negativity would exhibit the following scaling behavior:
	\begin{equation}
		\xi_{A_1:A_2} \sim \frac{c_{\rm eff}^{N}}{4} \ln L +\rm const,
	\end{equation}
	where $c_{\rm eff}^{N}$ represents the effective central charge of the 
	stationary state extracted using the log-negativity.

	\emph{Time evolution}.
	For noninteracting free-fermionic chains, owing to the simple relationship 
	between the eigenvalues of the reduced density matrix and two-point correlation 
	function, we can calculate the entanglement entropy more efficiently by 
	defining correlation matrix $W$ as

	\begin{equation}\label{eq:corr_mat_def}
		\begin{split}
			W_{nm}&=\tr \left( \rho \begin{bmatrix}
				f_n \\
				f_n^\dagger
			\end{bmatrix}
			\begin{bmatrix}
				f_m^\dagger & f_m
			\end{bmatrix}\right)\\
			&=\begin{bmatrix}
				\delta_{nm}-C_{nm} & F_{nm}\\
				\overline{F_{nm}} & C_{nm}
			\end{bmatrix},    
		\end{split}
	\end{equation}
	where $n,m=1,...,L$. The functions $C_{nm}$ and $F_{nm}$ are two-point 
	correlation functions, defined as $C_{nm}= \langle f_n^\dagger f_m\rangle$ and 
	$F_{nm}=\langle f_n f_m\rangle$. The expectation value $\langle \cdot \rangle$ 
	is taken with respect to the state of interest and the overbar represents 
	complex conjugation. The correlation matrix $W$ is a Hermitian matrix with 
	eigenvalues lying on the real interval $[0,1]$. Given the correlation matrix, 
	$S_{vN}$ can be evaluated as \cite{Vidal_2003,Peschel_2003,Ares_2015}
	\begin{equation}\label{eq:entropy_from_corrmat}
		S=-\frac{1}{2} \tr \left[(I-W) \ln (I-W)+W\ln W\right].
	\end{equation}
	At any finite time $t$, the correlation functions are given by $C_{nm}(t)= 
	\langle f_n^\dagger(t) f_m(t)\rangle$ and $ F_{nm}(t)=\langle f_n(t) 
	f_m(t)\rangle$, where the expectation values are taken with respect to the time 
	evolved state obtained under the operation of the postquench Hamiltonian. 
	The 
	two time-dependent correlation functions, $C_{nm}(t)$ and $F_{nm}(t)$, are 
	expressed as \cite{Krishnendu_2004}
	\begin{equation}
		\begin{aligned}
			\langle f_j(t) f_{j+l}(t) \rangle &= \frac{1}{2N} \sum_{n=0}^{N-1} e^{i k_n l} 
			\bigg(-\sin \left[2\lambda^f_{\alpha}(k_n) t\right] \sin \left(2 \delta 
			\theta_{k_n}\right) \\
			& \quad + i \bigg\{ \sin \left(2 \theta_{k_n}\right) \cos \left(2 \delta 
			\theta_{k_n}\right) \\
			& \qquad - \cos \left(2\lambda^f_{\alpha}(k_n) t\right) \sin \left(2 \delta 
			\theta_{k_n}\right) \cos \left(2 \theta_{k_n}\right) \bigg\} \bigg) \\
			\langle f_j^\dagger(t) f_{j+l}(t) \rangle &= \frac{1}{2N} \sum_{n=0}^{N-1} 
			e^{-i k_n l} \bigg\{ 1 - \cos \left(2 \theta_{k_n}\right) \cos \left(2 \delta 
			\theta_{k_n}\right) \\
			& \quad - \sin \left(2 \theta_{k_n}\right) \sin \left(2 \delta 
			\theta_{k_n}\right) \cos \left(2\lambda^f_{\alpha}(k_n) t\right) \bigg\},
		\end{aligned}
	\end{equation}
	where $\delta \theta_{k_n}=\theta^f_{k_n}-\theta^i_{k_n}$ is the difference
between pre- and postquench Bogoliubov angles defined in \cref{eq:bg_angle} and 
$\lambda^f_{\alpha}(k_n)$ is defined by \cref{eq:Ek_exp} for the postquench 
Hamiltonian. 
	
	In the limit $t \to \infty$, the time-dependent sine and cosine functions 
	become highly oscillatory and the respective sum would tend to zero. Using 
	this approximation, the expression for the correlation functions in the 
	long-time stationary state is given by
	\begin{align}\label{eq:correlators_sum_form}
		\begin{aligned}
			\langle  f_j f_{j+l} \rangle_{\text{st}}&= \frac{i}{2N} \sum_{n=0}^{N-1} 
			e^{i 
				k_n l} \sin \left(2 \theta_{k_n}\right)\cos\left(2 \delta \theta_{k_n}\right)\\
			\langle f_j ^\dagger  f_{j+l} \rangle_{\text{st}}&= \frac{1}{2N} 
			\sum_{n=0}^{N-1} e^{i k_n l}\left[1- \cos \left(2 
			\theta_{k_n}\right)\cos\left(2 \delta \theta_{k_n}\right)\right],
		\end{aligned}
	\end{align}
	where $\langle f_j ^\dagger  
	f_{j+l} \rangle_{\text{st}}$ and $\langle  f_j f_{j+l} \rangle_{\text{st}}$ are the two-point correlation and two-point anomalous correlation functions for the stationary state. We utilize the above two expressions to generate the 
	correlation matrix $W$, as 
	defined by \cref{eq:corr_mat_def}, and then apply 
	\cref{eq:entropy_from_corrmat} and \cref{eq:MI_def} to obtain $S_{vN}$ and 
	$I_{A_1:A_2}$, respectively, for the long-time stationary state.
	
	\textit{Numerical estimation of log-negativity}. Equation (\ref{eq:logneg_defn})
	suggests that to evaluate log-negativity, the crucial part is to perform the 
	partial transposition. Unfortunately, the partial transposition of a 
	fermionic 
	Gaussian state is not a Gaussian state
	\cite{Eisler_2015,Coser_2015}. This makes it difficult to efficiently calculate 
	log-negativity in noninteracting fermionic systems. However, by expressing the 
	partial transposition as a linear combination of two Gaussian operators, an 
	upper bound on log-negativity can be obtained and is expressed as 
	\cite{Coser_2015,Eisler_2015,Shapourian_2017,Eisert_2018,Rosz_2020,Sanku_2024,Herzog_2016}
	\begin{equation}\label{eq:lg_neg}
		\xi^u_{A_1:A_2}= \ln \tr \left( O_{+} O_{-}\right)^\frac{1}{2}+ \ln \sqrt{2},
	\end{equation}
	where the trace norm of $O_{+}$ is given by
	\begin{equation}
		\begin{aligned}\label{eq:O_plus}
			||O_{+}|| &= \tr \left( O_{+} O_{-}\right)^\frac{1}{2} \\
			&= \det \left[\left(\frac{\mathbb{I} + 
				i\Gamma_x}{2}\right)^{\frac{1}{2}}+\left(\frac{\mathbb{I} - 
				i\Gamma_x}{2}\right)^{\frac{1}{2}}\right] \\
			&\quad \times\det \left( \frac{\mathbb{I} -\Tilde{\Gamma}_1 
				\Tilde{\Gamma}_2}{2}\right),
		\end{aligned}
	\end{equation}
	with
	\begin{equation}\label{eq:gamma_x}
		\Gamma_x = i \left[1 - \left( 1 + i\tilde{\Gamma}_2 \right) \left( 1 - 
		\tilde{\Gamma}_1 \tilde{\Gamma}_2 \right)^{-1} \left( 1 + 
		i\tilde{\Gamma}_1 \right)\right]\,.
	\end{equation}
	The quantities $\Tilde{\Gamma}_1,\Tilde{\Gamma}_2$ in 
	\cref{eq:O_plus,eq:gamma_x} are defined as
	\begin{equation}\label{eq:Gamma_k}
		\Tilde{\Gamma}_k=\Tilde{M}_2 \Gamma_k \Tilde{M}_2\,, \quad \quad   
		M_2=\begin{bmatrix}
			\unity_L & 0\\
			0 & i \unity_L
		\end{bmatrix}\,,
	\end{equation}
	and the size of the subsystems $|A_1|=|A_2|=L$. The correlation matrices 
	$\Gamma_1$ and 
	$\Gamma_2$ are defined as
	\begin{equation}\label{eq:Gammas}
		\Gamma_1= \Gamma^A \quad \Gamma_2 = M_2 \Gamma_1 M_2 \quad 
		M_2=\begin{bmatrix}
			\unity_L & 0\\
			0 & -\unity_L
		\end{bmatrix},
	\end{equation}
	
	where $\Gamma^A$ is the correlation matrix corresponding to the subsystem $A$ 
	($A= A_1 \cup A_2$) and is defined as $\Gamma^A=2W-I$. With 
	\cref{eq:O_plus,eq:Gamma_k,eq:Gammas}, the upper bound of log-negativity can be 
	evaluated using \cref{eq:lg_neg}.
	
	\begin{figure}
		\centering
		\includegraphics[width=\linewidth]{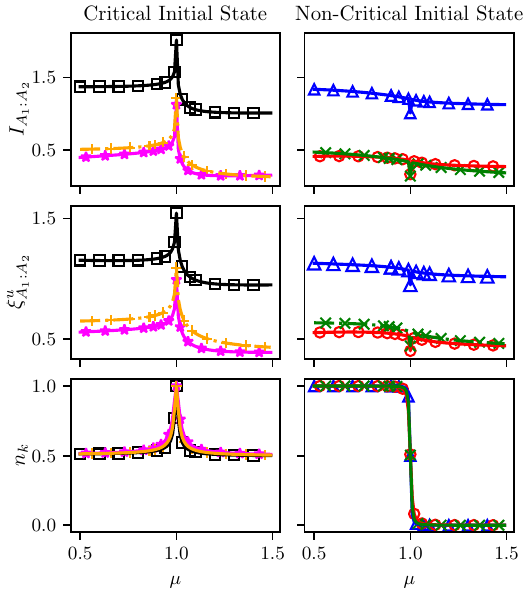}
		\caption{Top row: The mutual information between two 
		subsystems, $I_{A_1:A_2}$, of the stationary state as a function of the 
		postquench chemical potential $\mu_f=\mu$, for three values of 
		$\alpha$: 
		$\alpha = 0$ (black squares), $\alpha = 1$ (magenta stars), and $\alpha = 
		2$ (orange pluses). The prequench chemical 
		potential 
		$\mu_i$ is set at the critical point (left column), i.e., $\mu_i = 1$, and at a noncritical point (right column), i.e., $\mu_i = 1.5$. For noncritical initial state (right column), blue triangles
correspond to $\alpha=0$, red circles to $\alpha=1$ and green crosses to $\alpha=2$. Middle row: The log-negativity of the stationary state using 
		the same symbols and quench parameters as the top row. Bottom
		row: The occupation probability $n_k$ of Bogoliubov fermions at 
		the soft mode, calculated from Eq.~(\ref{eq:etakdagetak_exp}), with the same 
		symbols and quench parameters as in the top row. For all figures, the 
		quench protocol for the strength of the $p$-wave pairing interaction 
		$\Delta$ is chosen as $\Delta_i = -1$ and $\Delta_f = 1$.}
		\label{fig:stationary_mi_vs_mu}
	\end{figure}
	\section{Results}
	\subsection{Signature of criticality in stationary state}\label{sub:criticality}
	As we look for the signature of quantum criticality in the long-time 
	stationary state for different $\alpha$ values, we consider two types of 
	quench protocols which differ in the initial states. In the first protocol, 
	we consider the
	initial state as the ground state of the LRK Hamiltonian corresponding to the 
	parameter 
	values $\mu_i=1$ and $\Delta_i=-1$ and is henceforth referred to as the critical 
	state. The second protocol corresponds to the initial state as the ground state 
	of the
	LRK Hamiltonian corresponding to parameters $\mu_i=1.5$ and $\Delta_i=-1$, which 
	is henceforth referred to as the noncritical state. The final Hamiltonian parameters, 
	$\mu_f$ and $\Delta_f$, are varied to include both the critical and 
	noncritical 
	regimes. Note that it has already been established that mutual information of the
	ground state shows a peak at the critical point, namely, $\mu=\pm 1$ for large 
	$\alpha(=10)$, while only at $ \mu=1$ for $\alpha\leq 1$ 
	\cite{Lepori_2022}. Mutual information $I_{A_1:A_2}$ calculated for the
	stationary state for both 
	quench  protocols, i.e., $(\mu_i=1, \Delta_i=-1)\to (\mu_f=\mu, 
	\Delta_f=1)$ and 
	$(\mu_i=1.5, \Delta_i=-1)\to (\mu_f=\mu, \Delta_f=1)$, is plotted in 
	Fig.~\ref{fig:stationary_mi_vs_mu} (row 1). For the final quench parameters 
	corresponding to critical values, a peak in $I_{A_1:A_2}$ for the
	critical 
	initial state is clearly visible, 
	indicating the signature of the criticality in the stationary state 
	for all values of $\alpha$. In the other quench protocol with the
	noncritical initial state, $I_{A_1:A_2}$ does show a nonanalytic 
	behavior in the form of a sharp dip for $\mu_f=1$ for all values of $\alpha$. 
	To confirm that this behavior is indeed due to the quantum nature of correlations, 
	we plot log-negativity $\xi_{A_1:A_2}$ in \cref{eq:logneg_defn} in Fig. 
	\ref{fig:stationary_mi_vs_mu} (row 2) for the same quench protocols. The 
	similar peak and dip structure is unmistakably 
	seen in the behavior of $\xi_{A_1:A_2}$  whenever the final Hamiltonian is 
	critical.
	Such behavior of $I_{A_1:A_2}$ for the nearest-neighbor free-fermionic model 
	has earlier been 
	explained in terms of a soft mode for which the energy vanishes near criticality. 
	For a critical-to-critical quench, the soft mode does not get excited resulting 
	in a peak structure in $I_{A_1:A_2}$ while it heats up to infinite 
	temperature for a noncritical-to-critical quench displaying a dip in 
	$I_{A_1:A_2}$. This further manifests as an algebraic decay of fermionic 
	correlation for 
	critical-to-critical quench, akin to ground-state behavior, while for the 
	noncritical-to-critical quench, it decays exponentially, indicating a 
	noncritical behavior. However, such an 
	explanation falls short for the LRK model for $\alpha\leq 1$, 
	as the correlation decay pattern in the ground state is always algebraic.
	
	For this nonanalytic behavior of entanglement measure at $\mu_f=1$, we offer 
	an alternate explanation in terms of the mode occupation probability $n_k$ 
	for the soft mode ($k_c=\pi$) where the energy vanishes near criticality. Let us
	recall that $n_{k_c}=1/2$ would imply a contribution from multiple modes and 
	therefore leads to higher entanglement entropy for the union of two subsystems, 
	$S_{vN}^{A_1 \cup A_2}$. As 
	in $I_{A_1:A_2}$ in \cref{eq:MI_def}, $S_{vN}^{A_1 \cup A_2}$ is being 
	subtracted; this leads to a decrease in $I_{A_1:A_2}$. 
	In contrast, $n_{k_c}=1,0$ will lead to a reduction of 
	$S_{vN}^{A_1 \cup A_2}$ and, consequently, leads to an increment of 
	the mutual information. 
	
	The occupation probability, $n_k=\langle \eta_{k}^\dagger \eta_{k} \rangle$ of 
	the Bogoliubov modes, $\eta_k$, in the 
	stationary state of the LRK model is obtained as
	\begin{equation}\label{eq:etakdagetak_exp}
		\begin{aligned}
			&\langle \eta_{k}^\dagger \eta_{k} \rangle = \frac{1}{2} \left[1 - \cos 
			\left(2 \delta \theta_{k}\right)\right] \\
			&= \frac{1}{2} \left[1 - \frac{ \left(\mu_f + \cos k\right) 
				\left(\mu_i + \cos k\right) 
				+ \Delta_f \Delta_i g^2_\alpha \left(k\right)}
			{\lambda^f_{\alpha} 
				\left(k\right)\lambda^{i}_{\alpha}\left(k\right)}\right],
		\end{aligned}
	\end{equation}
	where $\lambda^{i}_{\alpha}\left(k\right) (\lambda^f_{\alpha}\left(k\right))$ 
	denotes the energy corresponding to the prequench (postquench) Hamiltonian. 
	From \cref{eq:etakdagetak_exp}, it can be shown that if we start with the 
	noncritical initial state, i.e., $\mu_i=1.5$, in the limit $k \to 
	\pi$, the 
	occupation probability $n_k$ is given by 
	\begin{align} 
		n_k = \begin{cases} 1 & \text{for } \mu < 1, \\ \frac{1}{2} & \text{for } \mu = 
			1,\\ 0 & \text{for } \mu > 1, 
		\end{cases} 
	\end{align} 
	as illustrated in Fig.~\ref{fig:stationary_mi_vs_mu}(row 3). 
	This explains the dips observed in 
	$I_{A_1,A_2}$ in the long-time stationary state for a noncritical-to-critical 
	quench. In contrast, for the quench protocol involving the critical state, 
	$n_k$ near the soft mode is 
	\begin{align} 
		n_k = 
		\begin{cases} 
			\frac{1}{2} & \text{for } \mu < 1, \\ 1 
			& \text{for } \mu = 1,\\ 
			\frac{1}{2} & \text{for } \mu > 1.
		\end{cases} 
	\end{align} 
	This explains the peak observed in $I_{A_1:A_2}$ for the 
	critical-to-critical quench protocol.
	
	{Going beyond the search for signatures of criticality in the 
	bipartite information-theoretic measures such as mutual information and 
	log-negativity, we now study the tripartite mutual information whose 
	negative, zero, and positive values indicate perfectly delocalized or 
	scrambled information, extensivity of mutual information, and redundancy of 
	information \cite{ROta_2016,Hosur_2016,Cacefo_2023}. For the completeness, the tripartite mutual information (TMI) 
	is defined as 
	\begin{equation}
		I_{A_1:A_2:A_3} = I_{A_1:A_2} + I_{A_1:A_3} -I_{A_1: A_2 \cup 
		A_3},
	\end{equation}
	where $I_{A_i: A_j}$ is the mutual information between $A_i$ and $A_j$ \cite{Cerf_1998}. 
	It is known that for 1D noncritical systems, $I_{A_1:A_2:A_3}$  approaches 
	zero in the
	large system limit \cite{Fagotti_2023}. For 
	the same quench protocols used in the study of mutual information, TMI for 
	the stationary state is presented in Fig. \ref{fig:trimi_vs_mu}. Similar to 
	the bipartite mutual information, the TMI for the stationary state obtained 
	post critical-to-critical quench also exhibits a peak at  $\mu=1$. In 
	contrast, a nonanalytical behavior of TMI  for the stationary state is 
	visible for noncritical-to-critical quench. This indicates that the 
	criticality associated with the stationary state can also be captured 
	through tripartite mutual information, which is a measure of nonlocal 
	correlations.
	\begin{figure}[h]
		\centering
		\includegraphics[width=\linewidth]{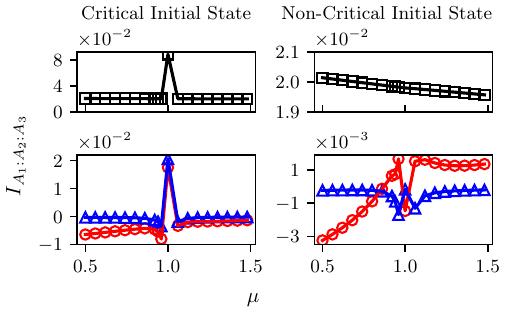}
		\caption{We plot the tripartite mutual information for different values 
		of post quench $\mu$ by fixing the prequench $\mu_i$ at $\mu_i=1$ (left 
		column) and $\mu_i=1.5$ (right column) for $\alpha=0$ (black squares), 
		$\alpha=1$ (red circles) and $\alpha=2$ (blue triangles). The 
		quench protocol for $\Delta$ is chosen to be the same as in  
		Fig.~\ref{fig:stationary_mi_vs_mu}}
		\label{fig:trimi_vs_mu}.
	\end{figure}
	Moreover, a negative TMI for $\mu<1$ shows the presence of nonlocal 
	information, but a redundancy of information for $\mu>1$ for $\alpha=1$. The 
	short-range pairing term, i.e., $\alpha=2$, displays zero TMI everywhere 
	except at the critical point, in agreement with the existing literature. An 
	all-to-all pairing term, however, has an entirely different behavior where 
	TMI is always positive, suggesting a redundancy of information, which peaks 
	at the critical point, $\mu=1$.
	  }
	
	\begin{figure*}[htbt]
		\centering
		\includegraphics{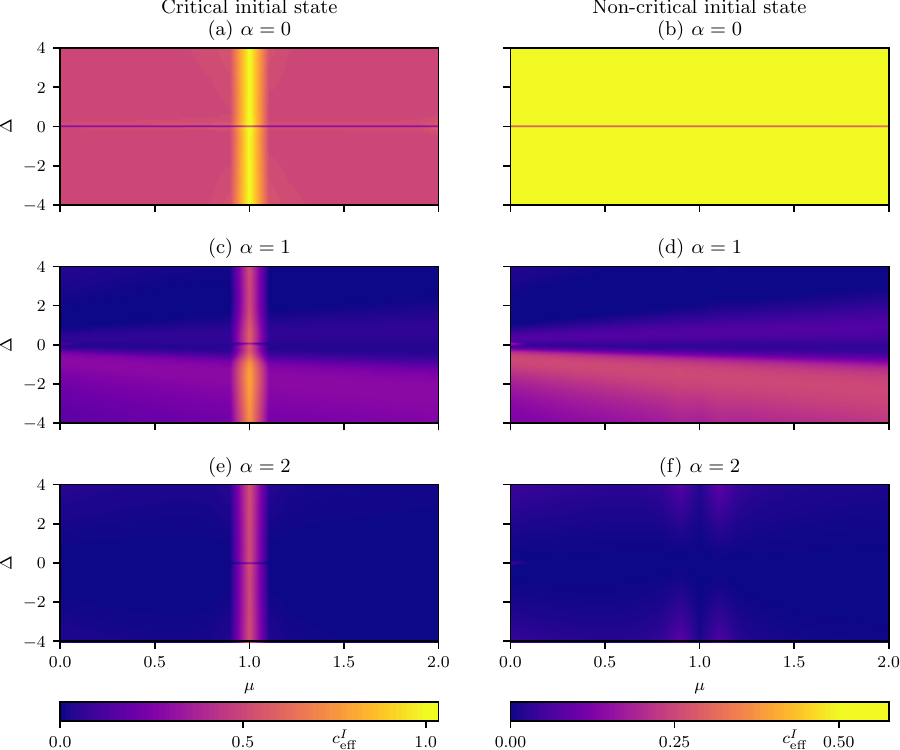}
		\caption{Phase plot of effective central charge $c_{\rm eff}^{I}$ 
		extracted from mutual information in the postquench
        $\Delta-\mu$ plane for both critical (left column) and noncritical 
        (right column) initial state for $\alpha=0$ (top row), 1 (middle 
        row) and 2 (bottom row). The parameter $\Delta_i$ is fixed at -1 
        for all the plots.}
		\label{fig:central_charge_mi_phase_plot}
	\end{figure*}
	
	\subsection{Signature of universality in stationary state}
	After establishing the nonanalytic behavior of mutual information (and 
	log-negativity) for the postquench stationary state at the critical point of the
	postquench 
	Hamiltonian, a natural question arises: Does the information of the effective 
	central charge, $c_{\rm eff}^I$ ($c_{\rm eff}^N$), 
	calculated from the scaling of $I_{A_1:A_2}$ ($\xi_{A_1:A_2}$), depend on the 
	quench protocol? And is this effective central charge universal, i.e., independent of quench protocols in
	$\Delta$; if yes, is the central charge of the ground state
	conserved? 
	To address these questions, we analyze both $c_{\rm eff}^I$ and $c_{\rm 
	eff}^N$ for both of the quench protocols discussed in section 
	\ref{sub:criticality}. Here, it is useful to 
	recall that central charge provides a powerful tool to understand the 
	underlying CFT that describes the particular universality class 
	\cite{Blote_1986}. For instance, the Ising universality class and the Luttinger 
	liquid universality class are described by a completely different CFT and 
	characterized by distinct central charges. The central charge gives a 
	measure of the number of degrees of freedom or the ``size" of the symmetry in the 
	theory \cite{Rattazzi_2011,Delfino_2021}. 
	
	Figure \ref{fig:central_charge_mi_phase_plot} displays $c^I_{\rm eff}$ as a 
	function of the postquench parameters $\mu_f$ and $\Delta_f$. The columns 
	of \cref{fig:central_charge_mi_phase_plot} represent $c^I_{\rm eff}$ for 
	quenches starting with two distinct initial states, while the rows correspond 
	to a varying range of $\alpha$. 
	Let us focus on the first column, i.e., 
	the quench from the critical state for different values of $\alpha$. For both 
	$\alpha=0$ 
	and $\alpha=2$, $c_{\rm eff}^{I}$ is independent of $\Delta$ quenches and 
	retains the same value as the ground state for a critical-to-critical quench 
	protocol.  For other values of $\mu_f$, 
	$c_{\rm eff}^{I}$ calculated from the stationary state takes the value of 
	noncritical ground state, i.e., $0.5$ for $\alpha=0$ and 0 for 
	$\alpha=2$ 
	\footnote{Note that $\Delta=0$ completely changes the Hamiltonian to a nearest 
		neighbour model with no pairing term with effective central charge 0 for $\mu \neq 
		1$.}. For $\alpha=1$, the effective central charge for the ground state at 
	the critical point 
	depends on the value of $\Delta$ and this character is retained in the 
	postquench 
	stationary state as well. Therefore, the universality can only be discussed in the 
	cases of $\alpha=0$ and $\alpha = 2$ from a critical-to-critical quench. It 
	is clearly seen 
	that $c_{\rm eff}^{I}$ calculated from the postquench stationary 
	state conserves the 
	central charge (see Table \ref{tab:ceff_gs}). This implies that the 
	universality class for the critical ground state can also be inferred from the
	stationary state.
	\begin{table}[h]
		\caption{Summary of effective central charge values calculated from 
		postquench 
			stationary state for critical-to-critical, noncritical-to-critical and 
			critical-to-noncritical 
			quenches of 
			LRK model for different values of $\alpha$.}\label{tab:ceff_ss}
		\begin{tabular}{|c|c|c|c|}
			\hline 
			~ & $\alpha=0$ & $0<\alpha \leq 1$ & $\alpha =2$\\ \hline 
			$\mu_i = 1, \mu_f=1$ & 1 & $c_\text{eff}(\alpha, \Delta)\neq 0$ & $\dfrac{1}{2}$\\ 
			[2mm]
			$\mu_i \neq 1, \mu_f=1$ & $\dfrac{1}{2}$ & $c_\text{eff}(\alpha, 
			\Delta) \neq 0$ 
			&  
			0\\ [2mm]
			$\mu_i = 1, \mu_f \neq 1$ & $\dfrac{1}{2}$ & $c_\text{eff}(\alpha, \Delta) \neq 0$ 
			&  
			0\\ [2mm]
			\hline
		\end{tabular}
	\end{table}
	
	Now for the second column of \cref{fig:central_charge_mi_phase_plot}, the
	universal characteristic of the $c_{\rm eff}^{I}$ of the stationary state is 
	reflected through the independence of $\Delta$ quenches for $\alpha=0$ and 
	$\alpha=2$, while $\alpha=1$ remains $\Delta$ dependent. Note that even 
	though the entanglement measures showed a non-analytic behavior at 	
	$\mu_f=1$ for the noncritical initial state, the effective central charge 
	does not 
	show any such transition for either $\alpha=0$ or 2. Therefore, despite the 
	nonanalytic behavior of entanglement measures at $\mu_f=1$, we do not 
	attribute it as a signature of criticality or a universal description in 
	terms of the underlying CFT. 

	Therefore, to summarize, the signature of 
	universality in terms of 
	effective central charge is seen in the postquench stationary state for a
	critical-to-critical quench. The effective central charges values in this case are 
	summarized in tabular form in \cref{tab:ceff_ss}, and are in agreement 
	with 
	the values in \cref{tab:ceff_gs} calculated for the ground state.

	Now, to investigate whether the long-range correlations in the 
	stationary state 
	are genuinely quantum in nature, we have evaluated the effective central charge 
	$c_{\rm eff}^{N}$ extracted from the finite-size scaling of the log-negativity for the stationary state. While the details are in 
	\cref{app:logneg}, the findings of the effective central charge behavior 
	and 
	therefore 
	universality remains the same as those calculated using mutual information. 
	
	\section{Summary}
	In summary, we studied the quench dynamics of the long-range Kitaev chain 
	to investigate the possible signatures of quantum criticality in the 
	long-time stationary state for different values of the exponent $\alpha$ 
	characterizing the range of $p$-wave pairing interaction. To probe the 
	criticality, we consider the quantum information-based measures such as 
	{bipartite, and tripartite} mutual information, and log-negativity. 
	Our results show that a peak emerges {at $\mu=1$ in bipartite 
	mutual information, tripartite mutual information}, and log-negativity for a 
	critical-to-critical quench for all values of $\alpha$ studied in this 
	work, exhibiting behavior similar to that of the critical ground state. 
	This clearly indicates the presence of criticality in the stationary state. 
	Importantly, {in contrast to the short-range $XY$ model,} the 
	appearance of this peak cannot be attributed to the change in decay 
	pattern of the fermionic correlation from exponential to algebraic at the 
	critical point, as the fermionic correlators decay algebraically even at 
	noncritical points for $\alpha \leq 1$. We argue that the peak in 
	bipartite mutual information arises due to the change in the occupation 
	probability of the Bogoliubov fermions in the soft mode at the critical 
	point. {To understand the longer range quantum information in 
	a postquench stationary state, we studied the TMI, which not only captures 
	the signature of criticality but also shows that quantum correlations are 
	delocalized for the $\mu<1$ case and have redundancy for $\mu>1$ for 
	$\alpha=1$. For pairing exponent $\alpha=2$, the mutual information is 
	extensive for all values of $\mu$ except $\mu=1$, while for $\alpha=0$, the 
	quantum correlations show redundancy.}
	
	We further show that for $\alpha = 2$, long-range correlations develop only 
	for the critical-to-critical quench. This result is obtained by analyzing 
	the effective central charges extracted from bipartite mutual information 
	and log-negativity scaling analysis. The effective central charge of the 
	corresponding stationary state matches the same for the critical ground 
	state, regardless of the quench in the strength of the $p$-wave pairing 
	interaction, $\Delta$. This match indicates that the universality class for 
	the critical ground state can also be inferred from the stationary state. 
	In contrast, for $\alpha = 0$, the effective central charge is nonzero for 
	all quench protocols, leading to long-range correlations, except when the 
	final Hamiltonian parameter $\Delta = 0$. When $\Delta \neq 0$, the 
	effective central charge is $1$ only for the critical-to-critical quench 
	and is $1/2$ for all other quench protocols, consistent with the ground 
	state. This consistency suggests that the stationary state for the critical-to-critical quench protocol belongs to the same universality class as the 
	ground state.
	
	For $\alpha = 1$, the stationary state, similar to the ground state, cannot 
	be described by any universality class, as the presence or absence of 
	long-range correlations depends on the $\Delta$ quenches. However, 
	long-range correlations develop for the critical-to-critical quench 
	protocol, regardless of the $\Delta$ quenches. 
	
	\section{Acknowledgment}
    S.P. would like to thank DST India for the Inspire Faculty Grant.
	
	\appendix
	\section{Overview of central charge for postquench stationary state 
		from log-negativity}\label{app:logneg}
\begin{figure*}[htbt]
		\centering
		\includegraphics{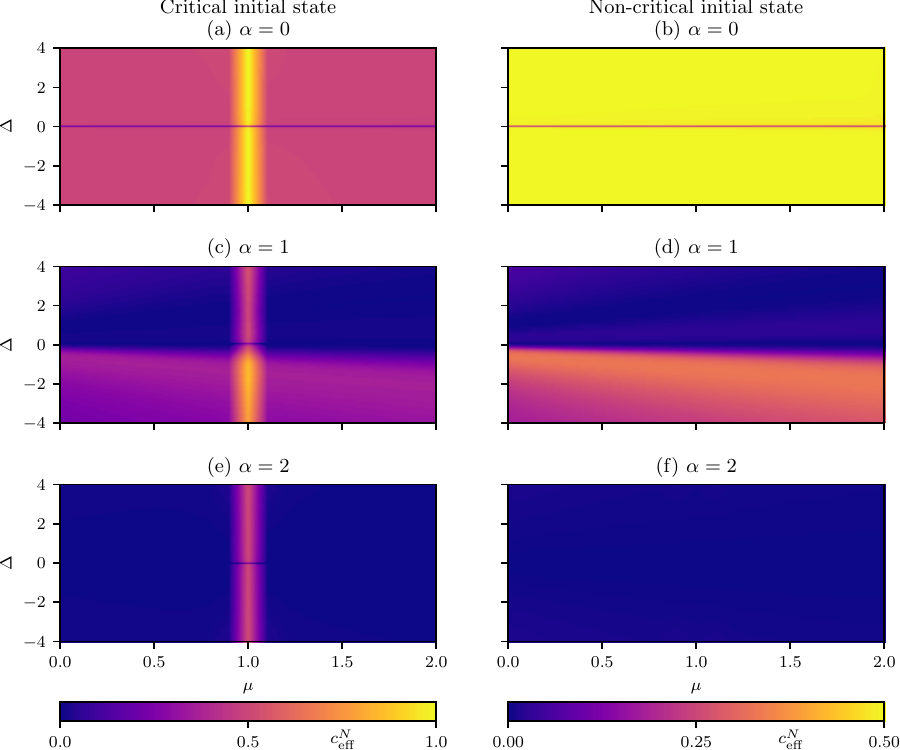}
		\caption{Phase plot of the effective central charge from 
        log-negativity $c_{\rm eff}^{N}$  for both critical (first column) and 
        noncritical (second column) initial state for $\alpha=0$ (first row), 
        1 (second row) and 2 (third row). The quench parameters are same as for 
        \cref{fig:central_charge_mi_phase_plot}.}
		\label{fig:central_charge_logneg_phase_plot}
	\end{figure*}

	Starting with the critical initial state, as shown in the top row, left column of 
	\cref{fig:central_charge_logneg_phase_plot}, we find that for
$\alpha = 0$, $c_{\rm eff}^{N}=1/2$ for $\mu_f \neq 1$ and $\Delta_f \neq 0$, 
while $c_{\rm eff}^{N}=1$ for $\mu_f = 1$ and $\Delta_f \neq 0$. In contrast, 
starting
with a noncritical initial state, as shown in the top row, right column of 
\cref{fig:central_charge_logneg_phase_plot}, $c_{\rm eff}^{N}=1/2$ for all values 
of $\mu_f$ and $\Delta_f$, except for $\Delta_f = 0$. This implies that for 
$\alpha = 0$, $c_{\rm eff}^{N} = c_{\rm eff}^{I}$ for all quench protocols. For 
$\alpha = 2$, similar to $c_{\rm eff}^{I}$, $c_{\rm eff}^{N}$ is nonzero and 
equal to $1/2$ only for a critical-to-critical quench protocol (with the exception 
of $\Delta_f = 0$, where $c_{\rm eff}^{N}=0$). For all other quench protocols, 
$c_{\rm eff}^{N}=0$. This confirms that the long-range correlations associated 
with a nonzero effective central charge for both $\alpha = 0$ and $\alpha = 2$ in 
the stationary state are strictly quantum in nature. On the other hand, for 
$\alpha = 1$, long-range correlations develop, irrespective of $\Delta$ quenches in 
the critical-to-critical quench protocol (see middle row, left column of 
\cref{fig:central_charge_logneg_phase_plot}). Interestingly, for this quench 
protocol, $c_{\rm eff}^{N}$ is slightly larger than $c_{\rm eff}^{I}$. In 
contrast, for the critical-to-noncritical quench protocol, the presence of 
long-range correlations in the stationary state depends on the $\Delta$ quenches. 
Specifically, if $\Delta_f \Delta_i \gtrsim 0$, both $c_{\rm eff}^{I}$ and $c_{\rm 
eff}^{N}$ are nonzero, whereas both are zero if $\Delta \Delta_0 \lesssim 0$. For 
all the quench protocols where $\Delta_f \Delta_i \gtrsim 0$, $c_{\rm eff}^{N} > 
c_{\rm eff}^{I}$.
	
	\bibliography{paper}
	
\end{document}